\documentclass[12pt]{article}
\newcommand{\be}{\begin{equation}}
\newcommand{\ee}{\end{equation}}
\usepackage{epsfig}
\begin{document}

\bigskip
\bigskip
\bigskip

\begin{center}
{\bf SOFT-HARD DECOMPOSITION IN QCD JET PHYSICS}
\end{center}

\bigskip

\begin{center}
{\bf I.M.~Dremin$^{(a)}$, A.V.~Leonidov$^{(a,b)}$ }
\end{center}

\begin{center}

\medskip

{\it (a) Theoretical Physics Department, P.N. Lebedev Physics Institute,\\ 119991 Leninsky pr. 53, Moscow, Russia}

\medskip

{\it (b) Institute of Theoretical and Experimental Physics\\ 117259 B. Cheremushkinskaya 25, Moscow, Russia}
\end{center}

\bigskip

\begin{abstract}
The paper generalizes the results on soft - hard decomposition of the
characteristics of QCD jets obtained in \cite{D05} by taking into account the
effects of fermions and running coupling constant.
\end{abstract}

\newpage


Physics of QCD jets is one of the most actively developing topics in application of Quantum Chromodynamics to phenomenology of multiparticle
production. The philosophy and practice of corresponding analysis is described
in \cite{DKMT91}, see also the reviews \cite{D94,KO97,DG01}. One of
the central issues in QCD applications to multiparticle production is distinguishing between the soft nonperturbative and (semi)hard perturbative
mechanisms of particle production \cite{DL95,L05}. Soft gluons play a very special role in applications of QCD parton model to multiparticle
production. This was a main motivation behind the analysis of \cite{D05}, where dynamics corresponding to perturbative evolution of QCD jets was
separated into soft and hard components according to the energy carried by an offspring parton. Such decomposition allows to make a separate analysis
of multiplicity distributions arising from soft and hard kinematical subspaces and suggest experimental observables allowing to verify the
theoretical picture. In \cite{D05} only gluon contribution was considered. In addition,
a simplifying assumption of a fixed (energy independent) QCD coupling constant
was made. In the present work we lift these two assumptions and study the soft-hard decomposition of QCD jets in a full picture that includes
fermions and the effects of running coupling constant.


The central object allowing to give a concise and parsimonious description of
the evolution of multiplicities of QCD jets is a generating function. It is
constructed from the probabilities of creating $n$ partons in a jet $P_n$:
\begin{equation}
G(z)=\sum_{n=0}^{\infty} P_n(1+z)^n.
\end{equation}
Introducing generating function allows to write a transparent system of integro-differential equations describing
the evolution of QCD jets:
\begin{eqnarray}
 \frac{d G_{G}}{dy} \; = \;\int_{0}^{1}dxK_{G}^{G}(x)\gamma _{0}^{2}[G_{G}(y+\ln x)G_{G} (y+\ln (1-x)) - G_{G}(y)] \nonumber \\
 \;+\;n_{f}\int_{0}^{1}dxK_{G}^{F}(x)\gamma _{0}^{2} [G_{F}(y+\ln x)G_{F}(y+\ln (1-x)) - G_{G}(y)], \label{eeg} \\
 \frac{d G_{F}}{dy} \; = \; \int_{0}^{1}dxK_{F}^{G}(x)\gamma _{0}^{2}[G_{G}(y+\ln x) G_{F}(y+\ln (1-x)) - G_{F}(y)] , \label{eeq}
\end{eqnarray}
where $y=\ln(p \, \theta /Q_0)$ is a parameter of jet evolution, $p$ is initial energy, $\theta$ is a jet opening
angle, $Q_0={\rm const}$, $\alpha_S$ is a strong coupling constant, $n_f$ is a number of quark flavors,
\begin{equation}
 \gamma_0^2 = \frac{ 2 N_c \alpha_S}{\pi}; \,\,\,\,\,\,\,\,\,
 \alpha_S(y)=\frac{2 \pi}{\beta_0 y} \left ( 1 - \frac{\beta_1 \ln 2y}{\beta_0^2 y} \right);
\end{equation}
\begin{equation}
 \beta_0  =  \frac{11 N_c - 2 N_f}{3};\;
 \,\,\, \beta_1 = \frac{17 N^2_c - N_f (5N_c+3 C_F)}{3}.
\end{equation}
The kernels of the evolution equations (\ref{eeg}) and (\ref{eeq}) read
\begin{eqnarray}
 K_{G}^{G}(x) \; = \; \frac {1}{x} - (1-x)[2-x(1-x)], \nonumber \\
 K_{G}^{F}(x) \; = \; \frac {1}{4N_c}[x^{2}+(1-x)^{2}], \nonumber \\
 K_{F}^{G}(x) \; = \; \frac {C_F}{N_c}\left[ \frac {1}{x}-1+\frac {x}{2}\right],   \label{K}
\end{eqnarray}
where $x$ and $1-x$ are the shares of energy carried by the created partons,
$N_c=3$ is a number of colors and $C_{F}=  (N_{c}^{2}-1)/2N_{c}=4/3$.

The main idea of \cite{D05} is to separate soft and hard contributions to jet multiplicity
$\langle n(y)\rangle=\langle n^S(y)\rangle+\langle n^H(y)\rangle$ by introducing a
cutoff $x_0 \ll 1$ in integration over
the phase space in the evolution equations (\ref{eeg}) and (\ref{eeq}). The
soft contribution $\langle n^S\rangle$ is that coming from integration in the interval
$[0,x_0]$ and the hard one $\langle n^H\rangle $ -- in the interval $[x_0,1]$.
Because of complete equivalence of two gluons emitted by their parent gluon,
the kernels of the equations are symmetrical for the substitution of $x$ by
$1-x$ in the initial formulation of the problem. It is, however, possible
to use this symmetry (see \cite{DKMT91}) and get the unsymmetrical kernels
(\ref{K}) at the expense of labeling one of the gluons as an offspring gluon.
Then the contributions $\langle n^S\rangle $ and $\langle n^H\rangle $ can be
easily separated. They differ in such formulation and correspond to mean
multiplicities of sets of jets with energies less than $x_0p$ and larger
than $x_0p$. There is no such problem for jets in
two-jet events of $e^+e^-$-annihilation because their energies are fixed from
the very beginning. However, subjets in these events, jets in three-jet events
or jets in hadronic reactions vary in energies. Therefore, it is important
to know the energy behavior of soft and hard sets of such jets. The above
proposal solves this problem.

In \cite{D05} only gluon evolution at fixed coupling constant was considered. Our aim in the
present paper is to generalize the calculations of \cite{D05} by taking into account the contributions of fermions and running coupling constant. The
calculations we have to perform mirror those in \cite{CDGNT00, DLN00}. The
only difference is the above-mentioned restriction in the integration over
the phase space in computing the soft contribution to jet multiplicity. Let us
remind that a convenient way of taking into account the effects of
running coupling \cite{D93} in the equations describing the evolution of the moments of jet multiplicity distributions with energy is to Taylor
expand the generating functions $G$ in the evolution equations in $y$, perform integration over $x$ and collect all contributions of the same order
in the right hand-- and left hand-- sides of the resulting equations.
Performing, in complete analogy with \cite{CDGNT00, DLN00}, this calculation
for the average soft multiplicities $\langle n^S_G\rangle $ and
$\langle n^S_F\rangle $ we get the following expressions for their second
derivatives:
\begin{eqnarray}
 \frac{d^2 \langle n_G^S\rangle }{dy^2}  =  \gamma_0 ^2 [\langle n_G\rangle x_0 +
 2(u\langle n_G\rangle -\langle n_G \rangle ^{'})w_1-(2u \, \langle n_G\rangle ^{'}
 -\langle n_G\rangle ^{''})w_2 +
 \frac{\langle n_G\rangle ^{'''}} {2}w_3                            \nonumber
\end{eqnarray}
$$
  + \frac{n_{f}}{4N_C}[(u\langle n_G\rangle -\langle n_G\rangle^{'}
   - 2u\langle n_F\rangle + 2\langle n_F\rangle ^{'})w_{4}+ 2(u\langle n_G\rangle ^{'}
   - 3u\langle n_F\rangle ^{'} + \langle n_F\rangle ^{''})w_5 + \langle n_F\rangle ^{'''}w_6]],
$$
\begin{eqnarray}
 r_0 \frac{d^2 \langle n^S_F\rangle }{dy^2} \; = \; \gamma_0^2
 [\langle n_G\rangle x_0
 +(u \, \langle n_G\rangle -\langle n_G\rangle ^{ '})w_7 +
 (2u \, \langle n_G\rangle ^{'} - \langle n_G\rangle ^{ ''})w_8
 - \frac{1}{2}\langle n_G\rangle ^{ '''}w_9  \nonumber\\
 \; + \; (u \, \langle n_G\rangle ^{ '}+u \, \langle n_F\rangle ^{'}-
 \langle n_F\rangle ^{ ''})w_{10}-\frac{1}{2}\langle n_F\rangle ^{ '''}w_{11}],   \label{nf}
\end{eqnarray}
where  $u=2B\gamma_0^2$, $B=\beta_0/8N_C$, $\beta_0=(11N_C-2N_f)/3$, $w_1(x_0) \ldots w_{11}(x_0)$ are the functions of the cutoff $x_0$ listed in
the Appendix.

The multiplicities of soft jets $\langle n_G^S\rangle $ and $\langle n_F^S\rangle$
can be obtained as the perturbative expansions in powers of $\gamma_0$:
\begin{eqnarray}\label{nn}
\langle n_G^S\rangle = \langle n_G\rangle [x_0-b_1\gamma_0-b_2\gamma_0^2 -b_3\gamma_0^3],\\ \langle n_F^S\rangle = \langle n_F\rangle
[x_0-d_1\gamma_0-d_2\gamma_0^2 -d_3\gamma_0^3].
\end{eqnarray}
\begin{equation}
 \gamma \; = \langle n_G\rangle ^{'} /\langle n_G\rangle \;= \gamma_0(1-a_1
 \gamma_0-a_2\gamma_0^2-a_3 \gamma_0^3)+O(\gamma_0^5), \label{gamma} \\
\end{equation}
\begin{equation}
 r \; =\langle n_G\rangle/\langle n_F\rangle \;=  r_0(1-r_1 \gamma_0-r_2
 \gamma_0^2-r_3\gamma_0^3)+O(\gamma_0^4),   \\  \label{r}
\end{equation}
$r_0=C_V/C_F = 9/4, \gamma_0^{'}\approx -B\gamma _0^3$.

The numerical values of $a_i, r_i$ are given in \cite{CDGNT00}.
Using these relations one easily gets the second derivative of
$\langle n_G^S\rangle$ as
\begin{equation}
\frac {\langle n_G^S\rangle ^{''}}{\langle n_G\rangle \gamma_0^2}= x_0-(b_1+(2a_1+B) x_0)\gamma_0-[b_2-(3B+2a_1)b_1-(2a_1B+a_1^2-2a_2)x_0]\gamma_0^2+
O(\gamma_0^3) \label{n''}
\end{equation}
and the similar expression for $\langle n_F^S\rangle ^{''}/\langle n_F\rangle $ with $b_i$ replaced by $d_i$. The terms of the same order of
$\gamma_0 $ must be equal on both sides of the equations. From this requirement one gets coefficients $b_i(x_0)$ and $d_i(x_0)$.
\begin{eqnarray}
 b_1 (x_0)=2w_1(x_0)+\frac{n_f}{4N_c}\frac {r_0-2}{r_0}w_4(x_0)-(2a_1+B) x_0, \\  \label{b1}
 b_2 (x_0)=- \left( 3(a_1+B)^2+ 2a_2 \right) x_0+2(a_1+B)w_1(x_0)-w_2(x_0) +\\
  \frac{n_f}{4N_c}[(a_1+B-\frac{2(a_1+B+r_1)} {r_0})w_4(x_0)-\frac{w_5(x_0)}{r_0}], \\ \label{b2}
 d_1(x_0)=w_7(x_0)-(2a_1+B)x_0 , \\ \label{d1}
 d_2(x_0)=-(3(a_1+B)^2+2a_2)x_0+(a_1+B-r_1)w_7(x_0)+w_8(x_0)+\frac{w_{10}(x_0)}{r_0}. \label{d2}
\end{eqnarray}
The numerical values of $b_i, d_i$ at $N_f=3$ for different $x_0$ are shown in Table 1:
\begin{center}
{\bf Table 1}

\medskip

\begin{tabular}{|c|c|c|c|c|}
  \hline
  $x_0$ & $b_1$ & $b_2$ & $d_1$ & $d_2$ \\ \hline
  0.1 & 0.095 & -0.417 & 0.004 & -0.42 \\ \hline
  0.2 & 0.03 & -0.447 & 0.003 & -0.748 \\ \hline
  0.3 & -0.181 & -0.53 & -0.014 & -1.276 \\ \hline
\end{tabular}
\end{center}
It is seen that the role of the second correction increases at lower values of $x_0$. The expressions for the terms of the third order of $\gamma _0$
are rather lengthy and we do not show them here. The results for gluodynamics are easily obtained if $N_F$ is put equal to zero in expressions for
$b_i$.

\begin{figure}[h]
\begin{center}
\epsfig{file=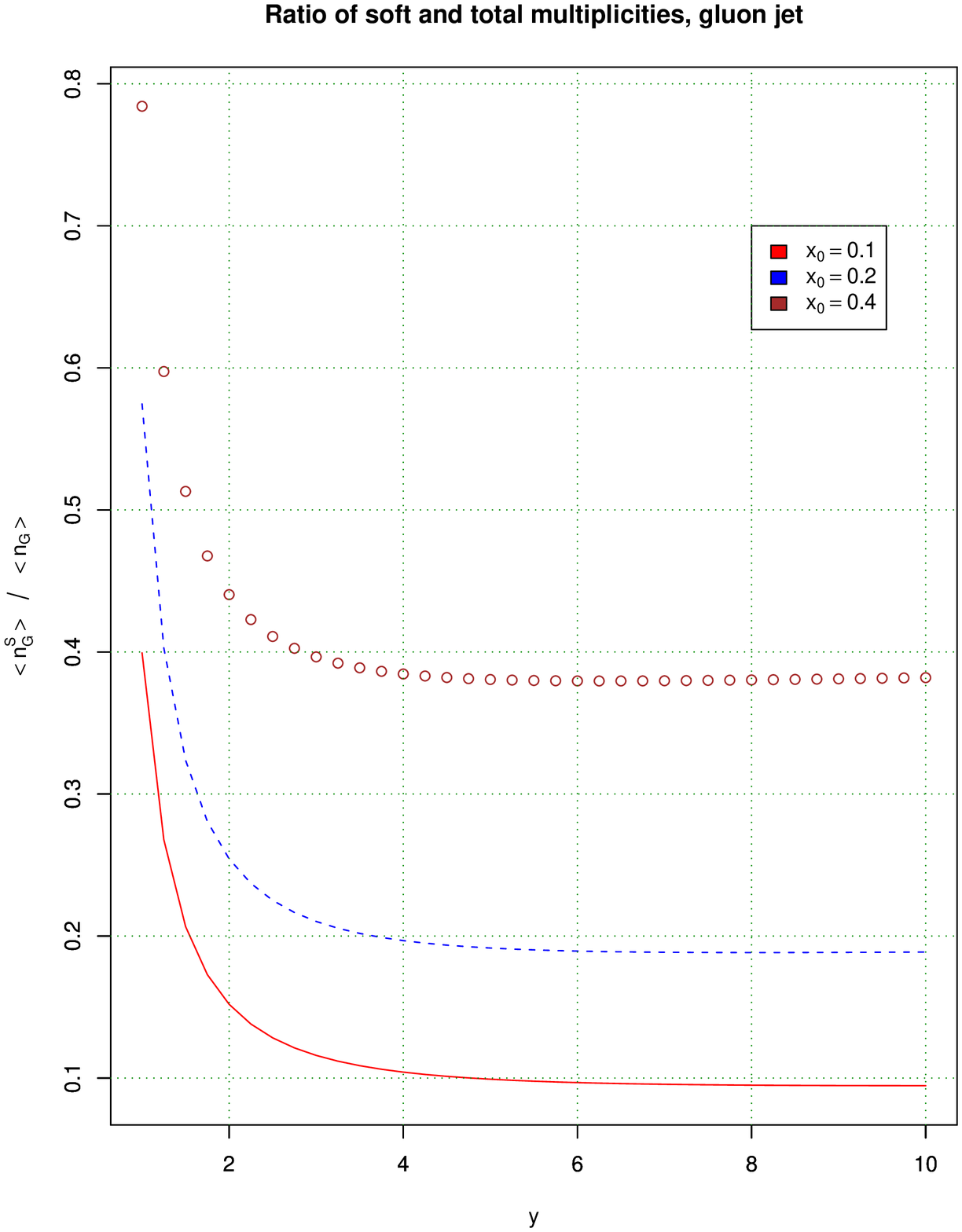,height=16cm,width=14cm}
\end{center}
\caption{ Energy dependence of the ratio of soft to total multiplicities in gluon jet $ \langle n^S_G \rangle /
\langle n_G \rangle $ for various cutoffs $x_0 \, = \, 0.1$ (solid line, red) $x_0 \, =  \, 0.2$ (dashed line,
blue) and $x_0= \, 0.4$ (dotted line, brown).} \label{ngns}
\end{figure}

The soft jet multiplicities behave asymptotically as the total multiplicities.
Some perturbative corrections in powers of $y^{-1/2}$ (or $\gamma _0$)
appear at lower energies. Soft multiplicities are generally proportional to
their softness parameter $x_0$ at small $x_0$ with terms like $x_0\ln x_0$
appearing in some $f_i$. All these conclusions can be tested in experiment.

Let us note that in the fixed coupling case considered in \cite{D05} the soft
jet multiplicities are proportional to the total multiplicity at all energies.
The proportionality factor depends only on $x_0$. It implies that the dependence
of their ratio at lower energies demonstrated in Fig. 1 is due to the running
property of the coupling strength in QCD.

This work has been supported in part by the RFBR grants 04-02-16445, 04-02-16880, 06-02-17051.\\

\newpage

\begin{center}
{\bf Appendix}
\end{center}

\begin{eqnarray}
 w_1(x_0) \; = \; \int^{x_0}_0 dx \, V(x) =\int^{x_0}_0 dx \, (1-\frac{3}{2}x +
x^2-\frac{x^3}{2}) =x_0-\frac{3}{4}x_0^2 + \frac{x_0^3}{3}-\frac{x_0^4}{8}, \nonumber \\
 w_2(x_0) \; = \; \int^{x_0}_0 dx \, [\frac{\ln(1-x)}{x}-2V(x) \, \ln x(1-x)], \nonumber \\
 w_3(x_0) \; = \; \int^{x_0}_0 dx \, [\frac{\ln^2(1-x)}{x}-2V(x) \, (\ln^2x+\ln^2(1-x))], \nonumber \\
 w_4(x_0) \; = \; x_0-x_0^2+\frac{2}{3}x_0^3, \nonumber \\
 w_5(x_0) \; = \; x_0(\ln x_0-1)-x_0^2(\ln x_0-\frac{1}{2})+2x_0^3(\frac{\ln x_0}{3}-\frac{1}{9}), \nonumber \\
 w_6(x_0) \; = \; \frac{2}{27}x_0^3(9\ln^2x_0-3\ln x_0+2)-\frac{x_0^2}{2}(\ln^2 x_0-\frac{1}{2}\ln x_0+\frac{1}{2}) +x_0(\ln^2 x_0 -2\ln x_0+2),
 \nonumber \\
 w_7(x_0) \; = \; \int^{x_0}_0 dx \, \Phi(x) =\int^{x_0}_0 (1-\frac{x}{2})dx=x_0-\frac{x_0^2}{4}, \nonumber \\
 w_8(x_0) \; = \; x_0(\ln x_0-1)-\frac{x_0^2}{4}(\ln x_0-\frac{1}{2}),\nonumber \\
 w_9(x_0) \; = \; x_0(\ln^2x_0-2\ln x_0+2) -\frac{x_0^2}{8}[2\ln^2 x_0-2\ln x_0 + 1 ], \nonumber \\
 w_{10}(x_0) \; = \; \int^{x_0}_0 dx \, [\Phi(x)-\frac{1}{x}]\,  \ln(1-x), \nonumber \\
 w_{11}(x_0) \; = \; \int^{x_0}_0 dx \, [\Phi(x)-\frac{1}{x}]\,  \ln^2(1-x) \, , \nonumber \\
\end{eqnarray}
The functions $V(x)$ and $\Phi(x)$ are defined in $w_1, w_7$.

\end{document}